# Deposition and characterization of TiZrV-Pd thin films by dc magnetron sputtering*

WANG Jie（王洁）, ZHANG Bo（张波）[1)], XU Yan-Hui（徐延辉）, WEI Wei（尉伟）, FAN Le（范乐）, PEI Xiang-Tao（裴香涛）, HONG Yuan-Zhi（洪远志）, WANG Yong（王勇）

National Synchrotron Radiation Laboratory，University of Science and Technology of China, HeFei, AnHui 230029 China

* Financially supported by the National Natural Science Funds of China (Grant No. 11205155) and Fundamental Research Funds for the Central Universities (WK2310000041).

1）corresponding author: zhbo@ustc.edu.cn；phone numbers: +8613615691450

## Abstract

TiZrV film is mainly applied in the ultra-high vacuum pipe of storage ring. Thin film coatings of palladium which was added onto the TiZrV film to increase the service life of nonevaporable getters and enhance pumping speed for $H_2$, was deposited on the inner face of stainless steel pipes by dc magnetron sputtering using argon gas as the sputtering gas. The TiZrV-Pd film properties were investigated by atomic force microscope (AFM), scanning electron microscope (SEM), X-ray photoelectron spectroscopy (XPS) and X-Ray Diffraction (XRD). The grain size of TiZrV and Pd film were about 0.42~1.3 nm and 8.5~18.25 nm respectively. It was found that the roughness of TiZrV films was small, about 2~4 nm, for Pd film it is large, about 17~19 nm. PP At. % of Pd in TiZrV/Pd films varied from 86.84 to 87.56 according to the XPS test results.

## 1. INTRODUCTION

Titanium-Zirconium-Vanadium (TiZrV) as one of nonevaporable getters (NEGs) [1-4] has been extensively studied during the past two decades for low secondary electron yield [5-7] and their sorption properties toward many gases such as hydrogen, oxygen, nitrogen, carbon monoxide and dioxide. The sorption of these gases except $H_2$ is not reversible and it causes a progressive contamination for TiZrV film [1, 8, 9]. Moreover, repeated air exposure–activation cycles progressively enrich the film with reactive gases, reducing its performance and shortening its operating life [10]. In order to enhance the lifetime of TiZrV film, the palladium overlayer is added on it. The contribution of the thin palladium film is particularly relevant to the pumping of hydrogen gas, due to its high sticking factor on palladium and the great sorption capacity of the underlying TiZrV getter.

Several researchers have done some experiments about absorbing behavior of TiZrV-Pd. M. Mura etc. [11, 12] designed an ion pump internally coated by TiZrV-Pd film, according to a technology that CERN (European Center for Nuclear Research) licensed them. The results showed that the pumping speed for $H_2$ was an order of magnitude higher, owing to the contribution of the getter. In addition, C. Benvenuti etc. [13] had studied the electron stimulated desorption and pumping speed measurements of TiZrV-Pd at CERN for particle accelerator applications. Nonetheless, it is necessary to have a further study into the effect of coating process and parameters such as discharge current, discharge voltage, working pressure etc. on the film structure, surface topography and grain size in dc magnetron sputtering process. Therefore, the aim of this paper is to study the problems.

## 2. EXPERIMENT

### 2.1. Coating equipment

A magnetron sputtering system was designed to coat TiZrV-Pd film onto the inner surface of a stainless steel pipe using argon as the sputtering gas. The schematic diagram is shown in Fig. 1. The chamber to be coated, 86 mm in diameter and 500 mm in length, is connected to two auxiliary chambers by Con-Flat flanges. The system is pumped by a turbo-molecular pump which is connected to the auxiliary chamber. The high purity argon intake is located on the auxiliary chamber and flow rate can be adjusted by a mass flow controlling system. There were two cathodes which were used one by one. And one cathodes was the twisting together Ti, Zr and V wires (2 mm in diameter) and the other was Pd wire (1 mm in diameter). Ten minutes was needed for this exchange of cathode wires. There is a ceramic insulation on the top of feedthrough. A ceramic cylinder is fixed at the end of the cathode to keep it insulated with the interior wall. In order to ensure the uniformity of film thickness, the cathode filament was fixed in the center of the stainless steel pipe using a bellows transmission mechanism (Fig. 1). Silicon substrates are mounted inside the chamber for evaluation of film thickness, grain size, morphology, and composition. The magnetic field is generated by a coaxial solenoid coil, which can generate magnetic fields up to 500 Gauss. A 3 KW DC power supply together with the coil are used to ignite and sustain the discharge.

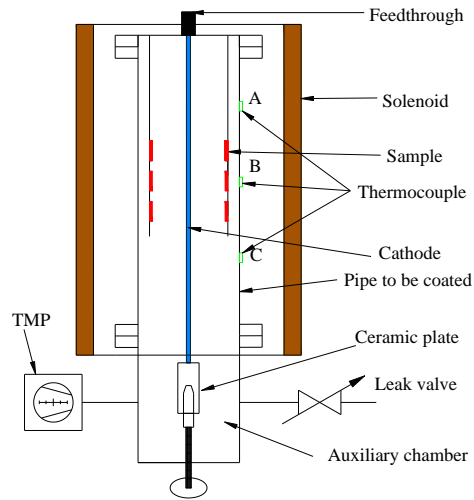

Fig. 1. Schematic diagram of TiZrV-Pd deposition system.

## 2.2. Magnetron sputtering process

Before deposition, Si substrates were ultrasonically degreased and cleaned in acetone and ethyl alcohol. Then they were dipped into the dilute HF solution, washed with deionized water and dried by purging with nitrogen gas. For the purpose of reducing the pollution from reactive gases, the pressure should be below $10^{-4}$ Pa before glow discharge.

Generally speaking, NEG-Pd film coating process was mainly divided into three steps. Firstly, the TiZrV wires cathode was power-on with selected film coating parameters. Secondly, after TiZrV film coating, nitrogen was passed into the pipeline before opening the flange. Thirdly, Pd wire cathode was installed as soon as possible, then film coating was started.

The roughness and porosity of TiZrV-Pd coatings increased with substrate temperature during deposition. Moreover, a rougher film surface could absorb more pollution gases such as carbon dioxide and oxygen. Consequently, the substrates were not intentionally heated during deposition. Three thermocouples were used to measure the temperature of the pipe wall, as shown in Fig. 1. On one hand, the temperature varied from 60 ℃ to 120 ℃ for TiZrV film coating with different sputtering current. On the other hand, the temperature varied from 30 ℃ to 100 ℃ for Pd film coating for the same reason. In Fig. 1, the temperature in point B was the highest owing to the worst heat dissipation in the central of solenoid which without cooling system. Then, the temperature in point A was higher than point C on account of the hot feedthrough during film coating process.

### 2.3. Characterization Method

Thickness were measured by use of a Sirion 200 Schottky field scanning electron microscope (SEM). Besides, material surface and internal compositional data were obtained with a Thermo ESCALAB 250 X-ray photoelectron spectroscopy (XPS). The spectrometer was equipped with a hemispherical analyzer, a monochromater, a beam spot size of 500 $\mu m$ and all XPS data was measured with Al Ka X-rays with ($h\nu$ =1486.6 eV) operated at 150 W and an analyzer at 45 degrees. Surface morphology was observed through Innova atomic force microscope (AFM) at room temperature. The crystal structure and the size of the crystallites was obtained by Rigaku TTR-III X-Ray Powder Diffraction (XRD). All XRD data was tested with Cu Kα X-rays which was used at 40 kV/200 mA. In addition, a diffractometer was used in a 2 $\theta/\theta$ mode, 2$\theta$ varying from 30 to 90 with a 0.02 step.

## 3. RESULTS AND DISCUSSION

### 3.1. Structural characterization

Bottom width of the diffraction peak is not easy to determine, so the actual work generally use half the peak width. Peak width is mainly affected by the following four factors: the uncertainty of wavelength, grain size, the position of instruments and samples, micro stress that exist within the scope of one or a few grain and keep the balance of internal stress. In the four factors, only the grain size was considered as the reason of peak width and other factors can be ignored. Under the assumption of a homogeneous single phase and roughly equiaxed crystal grains, the Scherrer formula is applied to determine the average dimension of the crystallites,

$$D = \frac{k\lambda}{W_{1/2} \cos\theta_B}$$

where $k$ is scherrer's constant which is depending on the crystal shape and related to grain size and distribution, often take 0.89. $\lambda$ is the wavelength of the source - i.e. 0.154 nm for copper $K\alpha$, $W_{1/2}$ is the angle in radians at half high width of diffraction peak for the certain crystal face wide, $\theta_B$ is the diffraction angle in degree, $D$ is the grain size.

According to the XRD test results, the grain size of TiZrV films was about 0.42（Sample #9-TiZrV）~ 1.3（Sample #2-TiZrV）nm as shown in Fig. 2. Meanwhile, the grain size was 8.5（Sample

#6- Pd）~ 18.25（#2- Pd）nm for Pd films as shown in Fig. 3. In the case of the same film coating conditions, the types of substrate have little effect on the grain size of Pd film. For instance, under the same film deposition condition, the grain size of Pd film which growed on TiZrV film was 14.9 nm, and the one on silicon (Si<111>) was 12.7 nm for sample #10-Pd. Furthermore, the film coating parameters of the samples were shown in Table 1. Possible causes are as follows. For the first Pd layer, different substrates will have an obvious impact on their arrangement. With the Pd atomic layers increases, this effect of substrate on Pd layer weakened little by little. While, on average, the effect of substrates on the grain size of Pd film is small, according to the test results which was shown in Table 2.

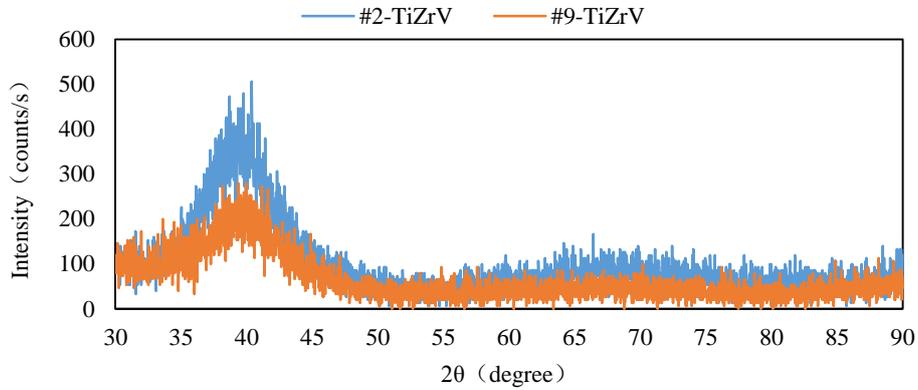

Fig. 2. X-ray diffraction of TiZrV films deposited at different film coating conditions.

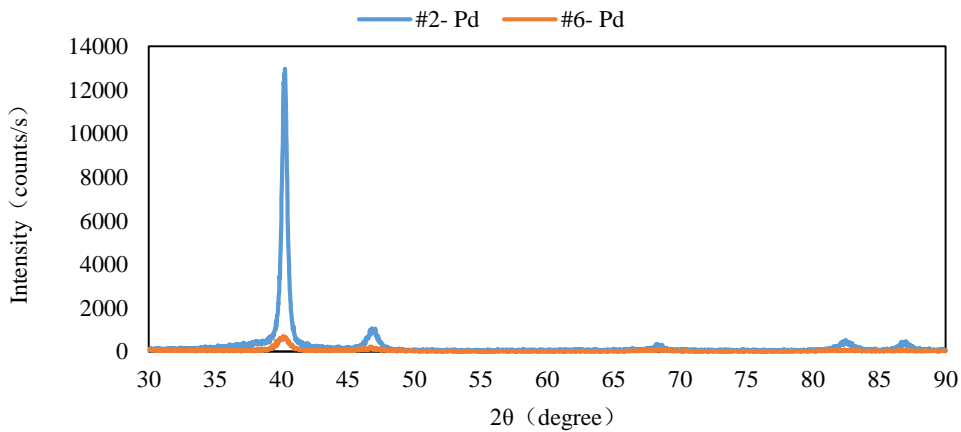

Fig. 3. X-ray diffraction of films comparison TiZrV/ Pd films deposited at different film coating conditions.

Table 1: Film coating parameters.

| Sample | Discharge Voltage/V | Discharge Current/A | Working Pressure/Pa | Magnetic Field/Gauss | Gas Flow/Sccm |
| --- | --- | --- | --- | --- | --- |

| Sample | | | | | |
|---|---|---|---|---|---|
| #1-TiZrV | 492-343 | 0.5 | 2.0 | 175 | 2.0 |
| #1- Pd | 530-523 | 0.04 | 2.0 | 175 | 2.0 |
| #2-TiZrV | 508-561 | 0.2 | 2.0 | 93 | 2.0 |
| #2- Pd | 444-453 | 0.02 | 2.0 | 235 | 2.0 |
| #3-TiZrV | 511-524 | 0.2 | 0.8 | 230 | 1.2 |
| #3- Pd | 430-440 | 0.02 | 2.0 | 206 | 2.0 |
| #4-TiZrV | 544-552 | 0.2 | 7.0 | 70 | 2.0 |
| #4- Pd | 443-440 | 0.02 | 2.0 | 64 | 2.0 |
| #5- Pd | 424-458 | 0.02 | 2.0 | 175 | 1.0 |
| #6- Pd | 329-354 | 0.2 | 20.0 | 123 | 2.0 |
| #7-TiZrV | 327-378 | 0.25 | 2.0 | 123 | 1.0 |
| #7-Pd | 437-548 | 0.04 | 5.0 | 175 | 2.0 |
| #8-Pd | 464-573 | 0.04 | 20.0 | 175 | 2.0 |
| #9-TiZrV | 400-347 | 0.25 | 2.0 | 123 | 4.0 |
| #10-Pd | 413-428 | 0.02 | 2.0 | 175 | 2.0 |
| #11-Pd | 432-440 | 0.02 | 2.0 | 232 | 2.0 |
| #12-Pd | 456-527 | 0.04 | 10.0 | 175 | 2.0 |

Table 2: The grain size of Pd film on different substrates according to XRD test results.

| Sample | grain size(Silicon-substrate) | grain size(TiZrV-substrate) | The thickness of Pd film (nm) |
|---|---|---|---|
| #10-Pd | 14.9 | 12.7 | 160 |
| #11-Pd | 17.4 | 17.3 | 220 |
| #8-Pd | 16.1 | 14.5 | 230 |
| #12-Pd | 12.3 | 13.1 | 270 |
| #7-Pd | 12.3 | 13.7 | 400 |

### 3.2. Surface morphology and section morphology

Fig. 5 was the section morphology images of samples #1-TiZrV,#1-Pd which manifests that the roughness of TiZrV films was small, mainly about 2~4 nm, for Pd films it is large, about 17~19 nm. What is more, the sputtering rates of samples #1-TiZrV,#1-Pd were 326 nm/h and 185 nm/h, respectively. On the basis of SEM surface topography measurement results: the grain size of TiZrV film was lower than Pd film. The surface topography of Pd films were shown in Fig. 5b and Fig. 5c, and the substrate was TiZrV film in Fig. 5c and silicon (Si<111>) in Fig. 5b.

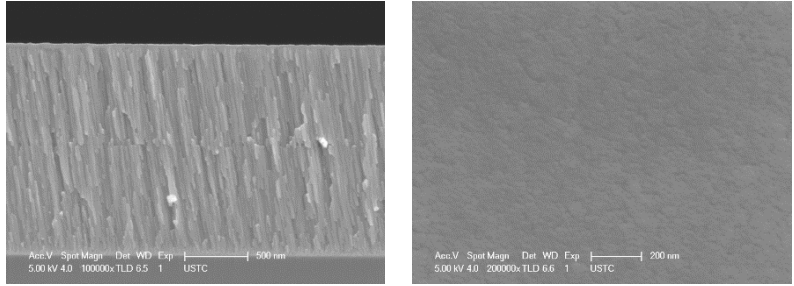

(a) #1-TiZrV

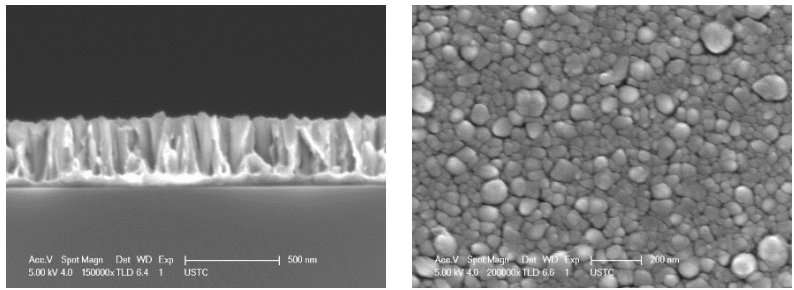

(b) #1- silicon/Pd

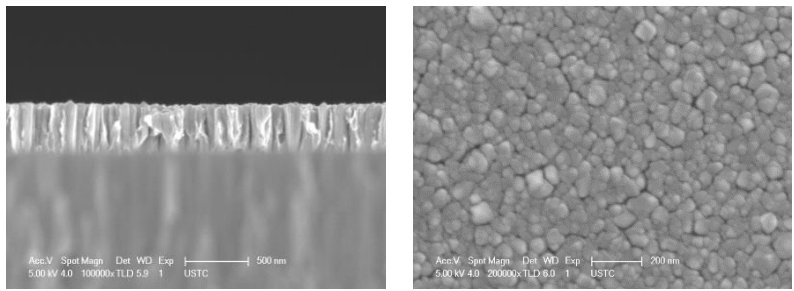

(c) #1-TiZrV /Pd

Fig. 5. Cross section morphology images (left) and surface topography images (right) of TiZrV film deposited on silicon by scanning electron microscopy (a), Pd film deposited on silicon (b), and Pd film deposited on TiZrV film. The film coating parameters were the same for Pd film in (b) and (c).

The surface of TiZrV film which was deposited on silicon was smooth and the largest degree of roughness was 3.6 nm with the scanning range of 5 $\mu m$ as shown in Fig. 6(a). While the surface of Pd film which was deposited on silicon was rough, compared with the one on TiZrV film, the degree of roughness was roughly 15.9 nm in Fig. 6(b). Furthermore, the roughness of Pd film which was deposited on TiZrV film was slightly larger than on silicon wafer, roughly 19.0 nm as shown in Fig. 6(c). Therefore, for the same silicon substrate, the roughness of TiZrV films were higher than Pd film under different film coating process. The main factor influencing the film surface roughness was the nature of the film rather than the substrates according to AFM test results.

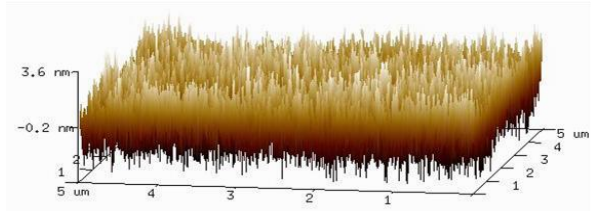

(a) #3-TiZrV

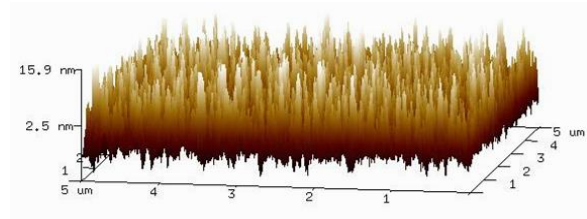

(b) #3-Pd

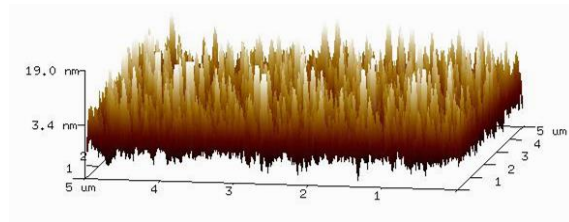

(c) #3-Pd

Fig. 6. AFM images of TiZrV film deposited on silicon (a), Pd film deposited on silicon (b), and Pd film deposited on TiZrV film which was shown in 6(c). The film coating parameters were the same for Pd film in (b) and (c).

Under the condition of discharge current 0.25 A, operating pressure 2.0 Pa, the magnetic field strength 123 Gauss, gas flow changed between 1.0 and 4.0 Sccm, AFM images of TiZrV film were shown in Fig. 7. Thus, it is obvious that the influence of gas flow to TiZrV film roughness was negligible.

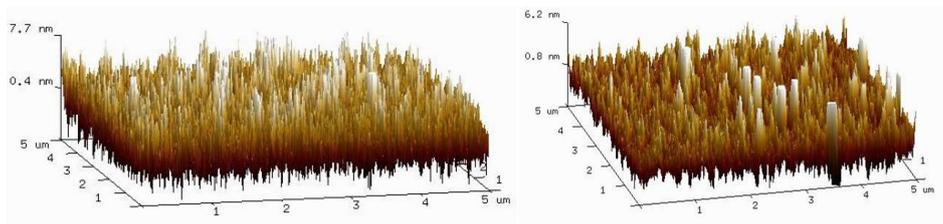

(a) #7-TiZrV    (b) #9-TiZrV

Fig. 7. (a) AFM images of TiZrV film deposited on silicon at 1.0 Sccm gas flow, (b) TiZrV film deposited on silicon at 4.0 Sccm gas flow, under the condition of discharge current 0.25 A, operating pressure 2.0 Pa, the magnetic field strength 123 Gauss.

### 3.3. Film composition analysis

After exposure to the atmosphere, a few nanometers of passivation layer which is mainly composed of oxide, carbide and nitride, formed on the surface. Fig. 8 illustrates that O had the highest atomic number percentage, and C was the second, N was the least in TiZrV film samples. Furthermore, V was mainly in the form of $V_2O_3$ because the binding energy of V was 516 eV. The peaks at around 459 eV, produced by the $TiO_2$, showed that air-exposed TiZrV film was partial oxidation. On account of a few nanometers depth with XPS detection, the surface element composition of the air-exposed TiZrV films usually were partial oxidation because of reactive metal Ti, Zr and V. Table 2 shows the ratio of the atomic number in TiZrV films based on the results of XRD.

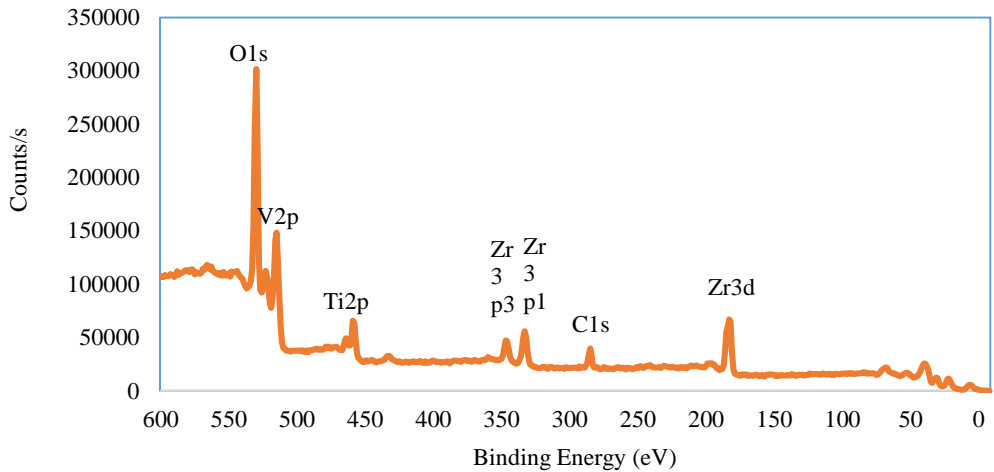

Fig. 8. X-ray photoelectron spectroscopy spectrogram of sample #4-TiZrV.

Table 2: The ratio of the atomic number in TiZrV films based on the results of X-ray photoelectron spectroscopy.

| Sample (TiZrV film) | Ti：Zr：V |
|---|---|
| #4-TiZrV | 2.4:2.8:4.8 |
| #7-TiZrV | 1.6:1.1:7.3 |
| #9-TiZrV | 1.4:1.1:7.5 |

The XPS of the air-exposed Pd film are shown in Fig. 9. The peaks at around 531 eV, produced by the surface oxide and carbon, showed that the air-exposed Pd film was contaminated in the process of transfer the sample. The peaks at 284 eV, produced by the carbon, showed that traces of organic carbon pollution, was detected on the unheated sample surface, very similar to that usually noticed for metal surfaces with different nature. In addition, PP At. % of Pd in Pd films varied from 86.84 to 87.56 based on XPS, shown in Table 3.

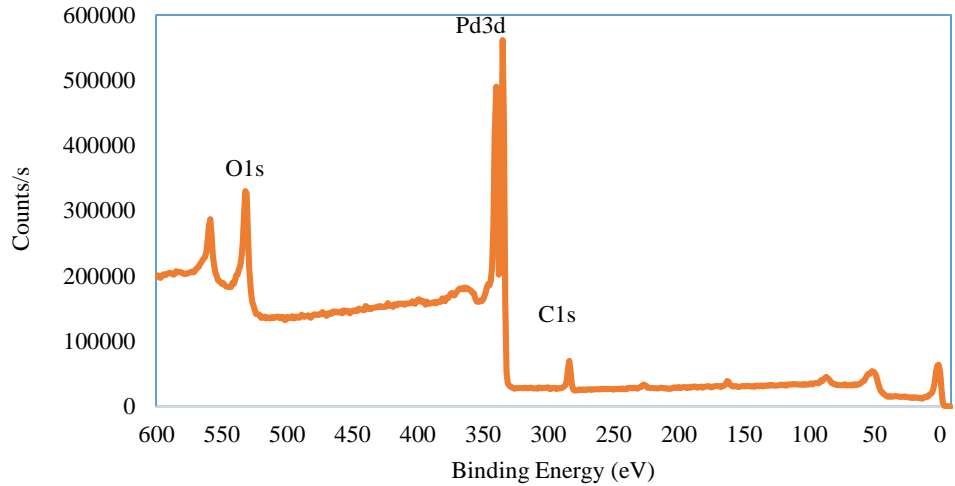

Fig. 9. X-ray photoelectron spectroscopy spectrogram of sample #4- Pd.

Table 3: PP At. % of Pd in TiZrV/Pd films based on the value of X-ray photoelectron spectroscopy.

| Sample (Pd film) | (PP At. %) |
| --- | --- |
| #5- Pd | 87.56 |
| #7- Pd | 86.84 |
| #8- Pd | 87.28 |

## 4. CONCLUSIONS

The combination of surface sensitive analysis techniques was used to study the deposition and characterization of TiZrV-Pd film coating by dc magnetron sputtering via investigation of surface composition and surface topography variations.

SEM and AFM test results consistently show that TiZrV films had a high consistency in thickness, and the thickness of Pd film fluctuated obviously. Moreover, the roughness of TiZrV films which were deposited on silicon were 3.6~5.0 nm with the scanning range of 5 $\mu m$ and Pd films which were deposited on TiZrV film were basically the same on silicon wafer, roughly 15.0~26.0 nm. Generally, the greater the working pressure, the greater the roughness of the film surface was, under the same gas flow, magnetic field strength, discharge current condition. Because the higher the working pressure, the more amount of residual gases were introduced into the film resulting in bigger roughness. On the other hand, in order to obtain high quality films, it was necessary to improve the vacuum degree of the system, since the residual gases were adsorbed by the surface of films in vacuum chamber during the process of film coating. The ratio of Ti, Zr, V atomic number varied between 2.4:2.8:4.8 and 1.1:1.6:7.3 in TiZrV films. In addition, PP At. % of Pd in TiZrV/Pd films varied from 86.84 to 87.56.